\documentclass{article} % For LaTeX2e
\usepackage{GEM_workshop_2024,times}

\usepackage{amsmath,amsfonts,bm}

\def\eqref#1{equation~\ref{#1}}
\def\1{\bm{1}}

\def\rd{{\textnormal{d}}}

\def\rvf{{\mathbf{f}}}

\def\rvm{{\mathbf{m}}}

\def\rvs{{\mathbf{s}}}

\def\rvv{{\mathbf{v}}}
\def\rvw{{\mathbf{w}}}
\def\rvx{{\mathbf{x}}}

\DeclareMathAlphabet{\mathsfit}{\encodingdefault}{\sfdefault}{m}{sl}
\SetMathAlphabet{\mathsfit}{bold}{\encodingdefault}{\sfdefault}{bx}{n}

\newcommand{\E}{\mathbb{E}}

\newcommand{\R}{\mathbb{R}}

\usepackage{hyperref}
\usepackage{url}

\usepackage{graphicx}
\usepackage{amsthm}
\usepackage{mathtools}
\usepackage{bbm}
\usepackage{subfigure}
\usepackage{wrapfig}
\usepackage{makecell}
\usepackage{algorithm}
\usepackage[noend]{algpseudocode}
\usepackage{caption}
\usepackage{booktabs}
\usepackage{amssymb}
\usepackage{colortbl}
\usepackage{pifont}% http://ctan.org/pkg/pifont
\newcommand{\cmark}{\ding{51}}%
\newcommand{\xmark}{\ding{55}}%

\definecolor{mydarkblue}{rgb}{0,0.08,0.45}

\hypersetup{
colorlinks=true,
urlcolor=mydarkblue,
linkcolor=mydarkblue,
citecolor=mydarkblue,
}

\let\emptyset\varnothing

\title{Fusing Neural and Physical: Augment Protein Conformation Sampling with Tractable Simulations}

\iclrfinalcopy

\author{Jiarui Lu\textsuperscript{1,2},  Zuobai Zhang\textsuperscript{1,2}, Bozitao Zhong\textsuperscript{1,2}, Chence Shi\textsuperscript{1,2}, Jian Tang\textsuperscript{1,3,4}
\\
  $^1$Mila - Qu\'ebec AI Institute, $^2$Universit\'e de Montr\'eal\\
  $^3$HEC Montr\'eal, $^4$CIFAR AI Chair \\
  \texttt{\{jiarui.lu, zuobai.zhang, bozitao.zhong, chence.shi\}@mila.quebec}, \\
  \texttt{jian.tang@hec.ca}
}

\begin{document}

\maketitle

\begin{abstract}

The protein dynamics are common and important for their biological functions and properties, the study of which usually involves time-consuming molecular dynamics (MD) simulations \textit{in silico}.  Recently, generative models has been leveraged as a surrogate sampler to obtain conformation ensembles with orders of magnitude faster and without requiring any simulation data (a "zero-shot" inference). However, being agnostic of the underlying energy landscape, the accuracy of such generative model may still be limited. In this work, we explore the few-shot setting of such pre-trained generative sampler which incorporates MD simulations in a tractable manner. Specifically, given a target protein of interest, we first acquire some seeding conformations from the pre-trained sampler followed by a number of physical simulations in parallel starting from these seeding samples. Then we fine-tuned the generative model using the simulation trajectories above to become a target-specific sampler. Experimental results demonstrated the superior performance of such few-shot conformation sampler at a tractable computational cost.
    
\end{abstract}
\section{Introduction}

Elucidating protein dynamics stands as a significant yet challenging problem during the study of protein functionality and regulation. Such macromolecule can go through transitions between multiple conformational states for different length or time scales. For example, the SARS-Cov-2 spike protein was found to have transitions between its open and closed states~\citep{gur2020conformational} for its function. Experimental instruments including  crystallographic B-factors and NMR spectroscopy can be leveraged to probe such dynamics but only available to a limited spatial and temporal scale. 

Traditionally, computational simulation methods especially molecular dynamics (MD) are used to simulate the dynamic behavior of these biological molecules. MD operates by simply evolving the Newtonian equation on the whole system of particles where the accelerations, or the gradients of velocity, are determined by a pre-specified force field (energy, or unnormalized density). To study the protein dynamics, the time scale of MD simulations can reach micro- to milli-seconds in order to completely capture the behaviors such as the transition from unfolding to folding~\citep{lindorff2011fast}. However, simulating protein systems in such time scales is very computationally intensive, which can take several hundreds of GPU days , depending on the size of the system.

With the emergence of accurate structure prediction models~\citep{jumper2021highly, baek2021accurate, lin2023evolutionary}, deep generative models are developed to serve as efficient surrogate for MD simulations, such as Str2Str~\citep{lu2024str2str}. Str2Str learns to explore the conformation space by training on protein structures from the Protein Data Bank (PDB) in an amortized way, and is ready to perform zero-shot conformation sampling for any unseen test proteins. The ability of direct sampling 
makes such sampler become orders of magnitude more efficient than traditional MD simulations. The inference of such diffusion sampler is based on successful transfer learning of the naturally occurred protein geometries. However, training solely on PDB data can make it fail to take care of the underlying energy landscape, which is, on the contrary, the fundamental consideration in MD simulation to generate Boltzmann-distributed ensembles. 

In this work, we aim to improve the diffusion sampler with tractable MD simulations via fine-tuning, forming a "few-shot"\footnote{Here we define "few-shot" in the sense that a limited quota of physical simulations for the test protein can be acquired, in contrast to the zero-shot setting where no simulation sample is present.} setting of Str2Str~\citep{lu2024str2str}. To be specific, given a target protein during inference, zero-shot samples are firstly generated from the pre-trained conformation sampler, followed by a number of "roll-out", i.e. short MD simulations respectively initialized by each sample, which is highly parallelizable and thus efficient. Then the production trajectories are leveraged to fine-tune the sampler \textit{ad hoc} for the target protein and reinforce the sampling accuracy. Experimental results demonstrate that such process significantly improve the quality of conformation sampling, at the cost of tractable computations compared to conventional simulations from scratch. 

\begin{figure}
    \centering
    \includegraphics[width=1\linewidth]{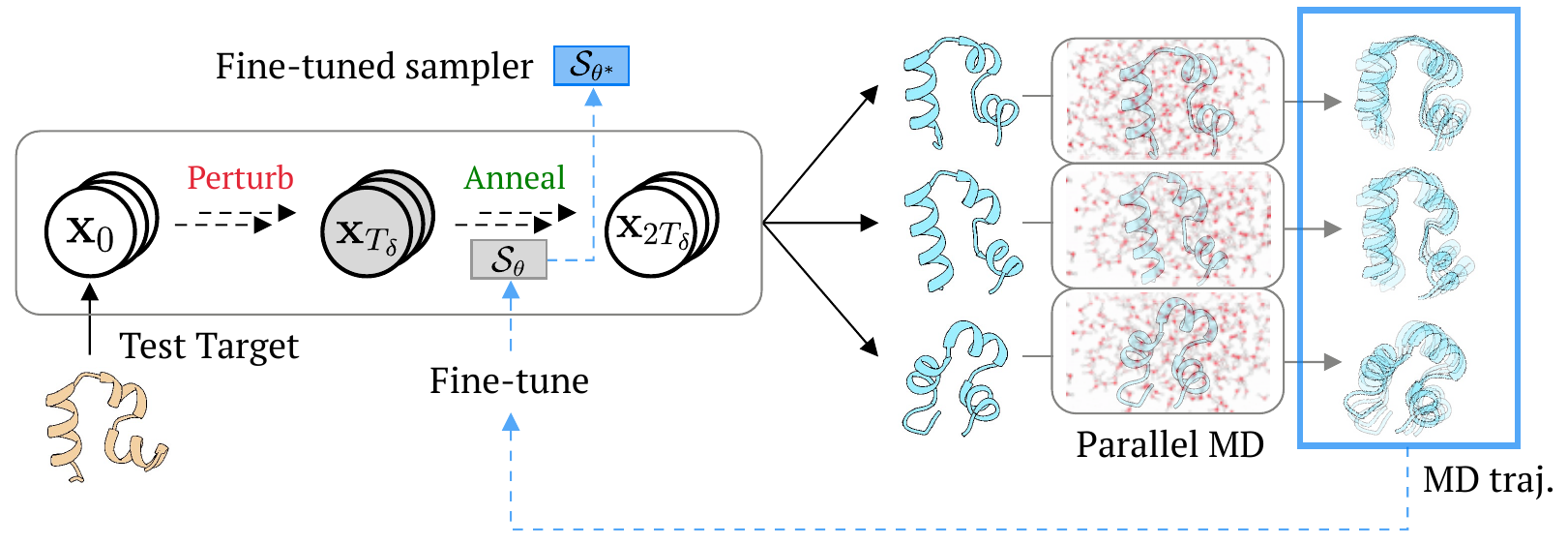}
    \caption{Illustrative diagram of fine-tuning the pre-trained diffusion sampler Str2Str~\citep{lu2024str2str}. Firstly, initial conformation samples of the target protein are generated from a pre-trained network $\mathcal{S}_\theta$ parameterized by $\theta$, followed by parallel MD simulations respectively for each sample. The production trajectories are leveraged to make a target-specific sampler $\mathcal{S}_{\theta^*}$ via fine-tuning. }
    \label{fig:ft}
\end{figure}
\section{Related work}

\paragraph{Protein conformation sampling}
Prior to this work, several methods have been proposed to sample protein conformations. Depending on the requirement of simulation data for training, these methods can be divided into two categories: \textbf{(1) Require simulation data.} Boltzmann generator~\citep{noe2019boltzmann} uses normalizing flow to approach the underlying Boltzmann distribution by learning from simulation data, which is one of the pioneer works among this research direction. 
\citet{arts2023two} instead adopts the denoising diffusion model to learn such distribution over the coarse-grained protein conformations.
idpGAN~\citep{janson2023direct} involves a conditional generative adversarial network trained on intrinsic disordered proteins.
DiG~\citep{zheng2023towards} trains a conditional diffusion model with both PDB and in-house simulation data and can sample conformations conditioned on input sequence. 
\textbf{(2) Require no simulation data.}
AlphaFold2-RAVE~\citep{vani2023alphafold2} injects stochasticity to the structure output by making modifications to the input channel of AlphaFold2~\citep{jumper2021highly}.
EigenFold~\citep{jing2023eigenfold} revisits the structure prediction (folding) problem in a generative view and is able to sample from the distribution over conformations given the input sequence. 
Str2Str~\citep{lu2024str2str} is an equivariant diffusion sampler trained solely on PDB data and formulates the conformation sampling in a structure-to-structure manner. 

\paragraph{\textit{De novo} design of protein structure}
Despite different purpose, protein structure design methods can also be relevant because of similar techniques used for modeling. \citet{anand2022protein} and \citet{shi2022protein} leverage the backbone architecture of AlphaFold2 but feed forward from the input conditions such as contact map. Afterwards, generative models are applied to learn from PDB and \textit{de novo} design novel backbone structures: ProtDiff~\citep{trippe2022diffusion}, FoldingDiff~\citep{wu2022protein},  RFDiffusion~\citep{watson2023novo}, Chroma~\citep{ingraham2023illuminating}, FrameDiff~\citep{yim2023se} and FoldFlow~\citep{bose2023se} to name a few. These sequence-agnostic generative models can design non-native backbone structures, based on which de novo sequences can further be designed by an inverse folding model, such as ProteinMPNN~\citep{dauparas2022robust}.
\section{Methodology}

\subsection{Preliminaries}
\paragraph{Protein structure.}
Protein structure consists of a sequence of amino acids (or residues) which respectively defines a set of atoms existing in three-dimension (3D) space. Formally, a protein with $L$ residues, while the residue $i$ ($1\le i \le L$) is composed of $n_i$ atoms, can be represented by the Euclidean coordinates of atoms $\rvx \in \R^{N\times 3}$, where $N = \sum_{i=1}^L n_i$ is the totally number of atoms. 

% by the Euclidean coordinates of the atoms $\rvx \in \R^{N\times 3}$ composing it, where $N$ is the number of heavy atoms. 

% Working with atoms directly is less efficient and thus we adopt the frame parameterizations~\cite{jumper2021highly}. In specific, for residue $i$, we define a tuple $T_i := (R_i, \vt_i)$ being the Euclidean transformation that determines the position and orientation. Here, $R_i \in \rm{SO(3)}$ is a $3\times 3$ rotation matrix while $\vv_i \in \R^3$ is a translation vector for the $i$-th residue.

\paragraph{Score-based generative models.}
Score-based generative models (SGMs)~\citep{song2020score} aim to model the data distribution via a diffusion process defined by the Itô stochastic differential equation (SDE):
\begin{equation}
\label{eq:sgm}
    \rd \rvx = \rvf(\rvx, t){\rd}t + g(t) {\rd}{\rvw},
\end{equation}
where continuous time $t\in [0, T]$, $\rvf(\rvx, t)  \in \R^n$ and $g(t)  \in \R$. The $\rvw \equiv \rvw(t) \in \R^n$ represents the standard Wiener process. Then, the corresponding backward SDE can be expressed as follow ~\citep{anderson1982reverse}:
\begin{equation}
\label{eq:bwd}
    \rd \rvx = [\rvf( \rvx, t) - g^2(t) \nabla_\rvx \log p_t(\rvx)] \rd t + g(t) {\rd}{\bar \rvw},
\end{equation}
where $\rd{t}$ is negative timestep and $\bar{\rvw}$ is the standard Wiener process with time $t$ flowing from $T \to 0$. 

\paragraph{Zero-shot conformation sampler.} Str2Str~\citep{lu2024str2str} leveraged SGMs to sample protein backbone conformation via a perturbation followed by an annealing process. Str2Str was trained solely with the protein structures in PDB and requires no simulation data during both training and inference. Str2Str demonstrated the effectiveness of SGMs for conformation sampling in a transfer learning manner, which shows promise when long MD simulation data is not readily available.

\paragraph{Molecular dynamics simulation.}

Molecular dynamics (MD) simulations amount to evolve the whole particle system through time $T^{\textrm{sim}}>0$ 
% \footnote{To avoid ambiguity with the time boundary $T$ in SGMs, the simulation time is denoted by $T^{\textrm{sim}}$ using superscript \textit{sim}.} 
by the physical dynamics ${d\rvx} = \rvv(\rvx,\tau){d\tau} $, where $\rvv \in \R^{N\times 3}$ is the velocity field. With some force field defining $U(\rvx)$ as potential energy, the velocity can be updated through some dynamics such as Newtonian equation\footnote{Sometimes referred to as Hamiltonian dynamics. Besides, stochastic dynamics can also be applied such as Langevin dynamics to update the velocity states.}: $d\rvv(\rvx, \tau) = \bar{\rvm}\nabla_\rvx U(\rvx) ~d\tau$, where $\bar{\rvm}$ is the inverse mass of atoms in the system. At each time step, the position and velocity can be updated by an integrator such as Verlet~\citep{verlet1967computer}, yielding a simulation trajectory. 
% Depending on the dynamics being studied, $T^{\textrm{sim}}$ has to reach a specific scale in order to sufficient reveal the equilibrium behaviors.

% ======================================================

\subsection{Neighborhood exploitation with short MD}

\paragraph{The dilemma in direct sampling and simulation.}
One disadvantage of Str2Str is that it has no prior knowledge about what do good conformations of the test protein look like during inference. On the other hand, MD simulations simply query a universal oracle (force field) at each time step, being generalizable and training-free, but have a hard time sufficiently exploring the conformation space within tractable computation for simulation. Previous methods, such as \citet{arts2023two}, proposed to train using the simulation trajectory of a target protein and sample like a generative model. Although achieving good performance, such method however requires expensive simulation data beforehand and cannot quickly generalize to other proteins beyond the training target, which in practice may have restricted potential utility. Here arises an interesting question: \textit{how to organically combine the advantages of both generative models and physical simulations?}

\begin{figure}
    \centering
    \includegraphics[width=0.85\linewidth]{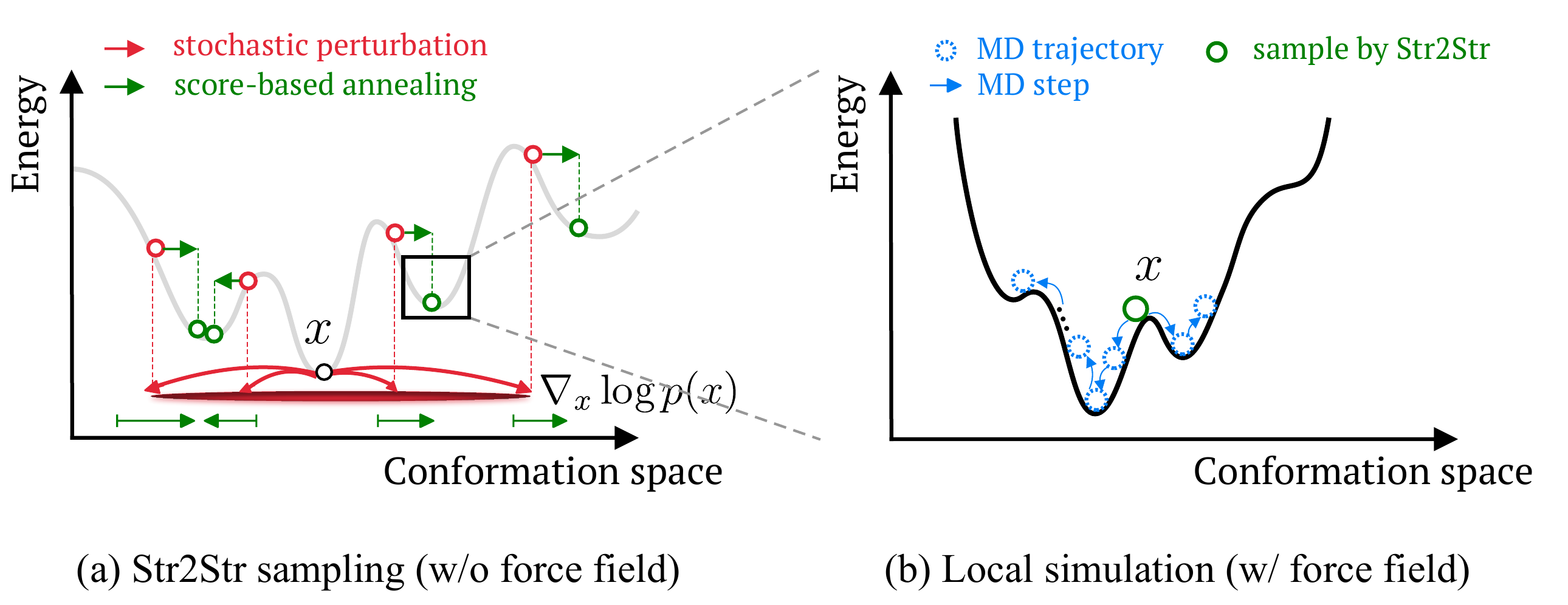}
    \caption{Illustration of the conformation sampling scenario of Str2Str. In (a), sampling is performed in two steps to obtain independent samples of the target protein, where the energy landscape (or information of the force field) is unknown and colored in {\color{gray}\textbf{gray}}. In (b), a hypothetical zoom-in neighborhood of a sample is shown. Due to the complex conformation landscape, the samples directly generated by Str2Str are probably not potential energy-optimal. Short MD simulation can be run to obtain locally equilibrated samples which can be used to fine-tune the pre-trained sampler.}
    \label{fig:illustration}
\end{figure}

\paragraph{Exploration-exploitation balance.}
To answer this question, we propose to sample protein ensembles via a two-stage sampling, or exploration-exploitation of the conformation space. Given a target protein of interest, we firstly leverage the Str2Str to sample a set of plausible conformations $X \equiv\{\rvx^{(i)}\}_{i=1}^m$, which can be seemingly good but not energy favorable. On top of that, a short MD simulation can be run for each sample $\rvx^{(i)}$ to produce (locally) equilibrated conformations. The resulting ensemble $X'$ will be the collection of conformations from each individual trajectory. The short MD simulations in the second stage exploits  the initial swarm  explored by the pre-trained diffusion sampler. The sampling with neighborhood exploitation (\textbf{Str2Str-NE}) enjoys (1) better exploration than conventional serial MD simulations under limited computation quota, as well as (2) better exploitation than diffusion sampler alone due to the awareness of force field.

\subsection{Fine-tuning of diffusion sampler}
The two-stage sampling proposed above can itself serve as a good sampler. Alternatively, given the refined conformations $X'$ as dataset, one can incorporate the supervision from the force field of MD simulation into the diffusion sampler Str2Str via fine-tuning (\textbf{Str2Str-FT}). In contrast to \citet{arts2023two} which involves expensive long MD trajectory for training, the acquisition of the production trajectories is very efficient: (1) The sampling (exploration) cost of diffusion sampler is negligible to MD simulations; (2) The computation of the local exploitation from short simulations can be readily paralleled according to the number of prior samples and enjoy perfect scalability. 
\begin{equation}
\label{eq:loss}
% \theta^* = \arg\min_\theta 
\mathcal{L}_{\textrm{FT}} = \E_{t\in [0, T]}  \,\E_{\rvx_0 \sim p(\rvx)} \left\{\omega(t, \rvx_0) \, \E_{\rvx_t\sim p(\rvx_t|\rvx_0)}\left[\|\rvs_\theta(\rvx_t,t) - \nabla_{\rvx_t} \log p_{t|0}(\rvx_t|\rvx_0) \|^2\right] \right\},
\end{equation}
where $\omega(t, \rvx_0) >0$ is a positive loss reweighting function depending both on time $t$ and sample $\rvx_0$, and $\rvx_t \sim p_{t|0}(\rvx_t|\rvx_0)$ is defined by the corresponding perturbation kernel. Among them, $\omega(t, \rvx_0)$ is a natural generalization of the importance sampling (IS) scheme for time in the training of diffusion model~\citep{song2020score}. The pseudo-code of the whole process is described in Algorithm \ref{algo:finetune}.

\begin{algorithm}
\caption{Fine-tuning of a pre-trained conformation sampler.}
\label{algo:finetune}
\begin{algorithmic}[1]
% \Procedure{Forward-backward}{}
\State\textbf{Require:} Target protein X, pre-trained diffusion sampler $\mathcal{S}_\theta$, MD simulator $\mathcal{M}$, simulation temperature $\beta$, simulation time $T^{\textrm{sim}}$.
\State $\{\rvx^{(i)}\}_{i=1}^m \sim \mathcal{S}_\theta(X)$ // Generate $m$ initial samples
\State $D_X = \emptyset$ // Initialize fine-tuning dataset
\For{$i=1$ to $m$}
    \State $\{\hat{\rvx}^{(i)}_j\}_{j=1}^{n_i} \sim \mathcal{M}(\rvx^{(i)}; \beta, T^{\textrm{sim}})$ // Run short MD simulation starting from $x_i$
    \State $D_X \gets D_X \cup \{\hat{\rvx}^{(i)}_j\}_{j=1}^{n_i}$ // Update dataset with the trajectory of size $n_i$
\EndFor
\State Update $\theta \rightarrow \theta^*$ based on $D_X$ by training on Eq. \ref{eq:loss}.
\State \textbf{return} $s_{\theta^*}$
% \EndProcedure
\end{algorithmic}
\end{algorithm}

\section{Experiments}

\paragraph{Setup.}
To validate the effectiveness of the proposed fine-tuning pipeline, we evaluate the sampling performance on the fast-folding protein set~\citep{lindorff2011fast}. For baselines, we compare the fine-tuned Str2Str (namely Str2Str-FT) against: MSA subsampling~\citep{del2022sampling}, EigenFold~\citep{jing2023eigenfold}, idpGAN~\citep{janson2023direct}, and Str2Str (pre-trained only)~\citep{lu2024str2str}. The size of sampled ensembles is aligned to be 1,000 to make fair comparison. We follow the evaluation metrics used in \citet{lu2024str2str} and list them as below: (1) Validity: \textit{Val-Clash}, \textit{Val-Bond} evaluates whether the conformations obey basic geometric constraints w.r.t. steric clash and bonding association. These metrics are the higher the better ($\uparrow$). (2) Distance: \textit{JS-PwD}, \textit{JS-TIC}, \textit{JS-Rg} assess the distance between the model distribution and reference distribution, which includes the pairwise distance (PwD), time-lagged independent components (TIC)~\citep{naritomi2011slow, perez2013identification} and radius of gyration (Rg), which are the lower the better ($\downarrow$). 

\paragraph{Benchmark results.}
In this study, we limit the few-shot budget of MD production simulations to be \textit{100ns} in total, which is computationally tractable and lasts only several GPU hours. Under this budget, we firstly sample for each target 100 conformations from a pre-trained Str2Str followed by 100 short simulations per 1 ns long. As shown in Table \ref{tb:science2011}, both the sampler with neighborhood exploitation (Str2Str-NE) and the fine-tuned version (Str2Str-FT) achieve the state-of-the-art performance across all three distance metrics while keeping good validity.

\begin{table}[!ht]
    \centering
    % \scriptsize
    % \small
    \caption{Benchmark results of conformation sampling methods on fast folding proteins~\citep{lindorff2011fast}. Metrics are averaged across all fast-folding targets for each method. The results of Str2Str is reproduced by running the open-source code, while the results of other baselines are taken from \citet{lu2024str2str}. The best result for each section is \textbf{bolded}. }
    \label{tb:science2011}
    \begin{tabular}{c|ccccc}%{c|c|c|c|c|c|c}
    \toprule[1pt]
        Methods & Val-Clash($\uparrow$) & Val-Bond($\uparrow$) & {JS-PwD}($\downarrow$) & {JS-TIC}($\downarrow$) &  JS-Rg ($\downarrow$)   \\ 
        \midrule
        MSA subsampling & \textbf{0.999} & 0.997 & 0.634 & 0.624 & 0.656 \\
        EigenFold & 0.812 & 0.874 & 0.530 & 0.497 & 0.666  \\
        idpGAN & 0.960 & 0.032 & 0.480 & 0.517 & 0.661\\
        Str2Str & 0.977 & 0.982 & 0.348 & 0.400 & 0.365 \\
        
        \midrule
        Str2Str-NE (ours) & 0.990 & \textbf{1.000} & \textbf{0.294}  & 0.369 & \textbf{0.331} \\
        Str2Str-FT (ours) & 0.966 & 0.948 & 0.303 & \textbf{0.366} & 0.350 \\

        % ablations 
        % \method (SDE) & 0.980 & 0.984 & \textbf{0.355} & 0.379 & \textbf{0.370} & \textbf{0.137} & \textbf{0.198} \\
        
        % {\color{brown} Reference 100ns} & 1.000 & 1.000 & 0.458 & 0.491 & 0.445 & 0.227 & 0.379 \\
        
        % {\color{brown}Reference 1us} & 1.000 & 1.000 & 0.317 & 0.392 & 0.308 & 0.206 & 0.339 \\
        % {\color{brown}Reference 10us} & 1.000 & 1.000 & 0.236 & 0.316 & 0.236 & 0.144 & 0.243 \\
        % {\color{brown}Reference 100us} & 0.997 & 1.000 & 0.130 & 0.155 & 0.129 & 0.063 & 0.102 \\
        % {\color{brown}Reference Full} & 0.997 & 1.000 & \cellcolor{black!15} 0.000 & \cellcolor{black!15}0.000 & \cellcolor{black!15}0.000 & \cellcolor{black!15}0.000 & \cellcolor{black!15}0.000 \\
       
        \bottomrule[1pt]
    \end{tabular}
\end{table}

\paragraph{Ablation study.}
To better demonstrate the effectiveness of the proposed Str2Str-FT, we also consider ablation experiments. The base model is the (1) Str2Str-FT sampler which is fine-tuned by samples from Str2Str-NE. We (2) replace the fine-tune data with conventional MD simulation of 100ns; and (3)~(4) ablate the pre-training from both settings. The results are shown in Table \ref{tb:abl}.

\begin{table}[!ht]
    \centering
    % \scriptsize
    \small
    \caption{Ablation results on different strategies of training. "NE" indicates the training set is from the samples of Str2Str-NE introduced above while "MD" from the conventional MD simulations. "FT" means model parameters are initialized from the PDB pre-trained checkpoint (i.e. Str2Str).}
    \label{tb:abl}
    \begin{tabular}{cccc|ccccc}%{c|c|c|c|c|c|c}
    \toprule[1pt]
        Settings & PDB & MD & NE & Val-Clash($\uparrow$) & Val-Bond($\uparrow$) & {JS-PwD}($\downarrow$) & {JS-TIC}($\downarrow$) &  JS-Rg ($\downarrow$) \\ 
        \midrule
        
        % new baselines
        FT on NE data & \cmark &\xmark & \cmark  & 0.966 & 0.948 & \textbf{0.303} & \textbf{0.366} & \textbf{0.350} \\
        FT on MD data &\cmark &\cmark & \xmark & \textbf{0.999} & \textbf{0.961} & 0.538 & 0.570 & 0.544  \\
        NE data  &\xmark &\xmark & \cmark & 0.227 & 0.562 & 0.398 &  0.406 & 0.473  \\
        MD data &\xmark &\cmark & \xmark  & 0.835 & 0.632 & 0.549 & 0.534 & 0.603  \\

        % \method (SDE) & 0.980 & 0.984 & \textbf{0.355} & 0.379 & \textbf{0.370} & \textbf{0.137} & \textbf{0.198} \\
       
        \bottomrule[1pt]
    \end{tabular}
\end{table}
\section{Conclusion}

In this work, we explore the few-shot settings of protein conformation sampling by combining score-based generative models and physical MD simulations. 
Generative models explore the conformation space efficiently while the physical force field refine the unreasonable samples, achieving good exploration-exploitation balance.
The fine-tuned diffusion sampler has shown significant improvement on conformation sampling, which exemplified the domain adaption of Str2Str with efficient short simulations. 
Future works can focus on the study of more effective fine-tuning strategies to enable sampling approximately from Boltzmann distribution.


\subsubsection*{Acknowledgments}
This project is supported by Twitter, Intel, the Natural Sciences and Engineering Research Council (NSERC) Discovery Grant, the Canada CIFAR AI Chair Program, Samsung Electronics Co., Ltd., Amazon Faculty Research Award, Tencent AI Lab Rhino-Bird Gift Fund, an NRC Collaborative R\&D Project (AI4D-CORE-06) as well as the IVADO Fundamental Research Project grant PRF-2019-3583139727.

\bibliography{iclr2024_conference}

\begin{thebibliography}{39}
\providecommand{\natexlab}[1]{#1}
\providecommand{\url}[1]{\texttt{#1}}
\expandafter\ifx\csname urlstyle\endcsname\relax
  \providecommand{\doi}[1]{doi: #1}\else
  \providecommand{\doi}{doi: \begingroup \urlstyle{rm}\Url}\fi

\bibitem[Anand \& Achim(2022)Anand and Achim]{anand2022protein}
Namrata Anand and Tudor Achim.
\newblock Protein structure and sequence generation with equivariant denoising diffusion probabilistic models.
\newblock \emph{arXiv preprint arXiv:2205.15019}, 2022.

\bibitem[Anderson(1982)]{anderson1982reverse}
Brian~DO Anderson.
\newblock Reverse-time diffusion equation models.
\newblock \emph{Stochastic Processes and their Applications}, 12\penalty0 (3):\penalty0 313--326, 1982.

\bibitem[Arts et~al.(2023)Arts, Satorras, Huang, Zuegner, Federici, Clementi, No{\'e}, Pinsler, and Berg]{arts2023two}
Marloes Arts, Victor~Garcia Satorras, Chin-Wei Huang, Daniel Zuegner, Marco Federici, Cecilia Clementi, Frank No{\'e}, Robert Pinsler, and Rianne van~den Berg.
\newblock Two for one: Diffusion models and force fields for coarse-grained molecular dynamics.
\newblock \emph{arXiv preprint arXiv:2302.00600}, 2023.

\bibitem[Baek et~al.(2021)Baek, DiMaio, Anishchenko, Dauparas, Ovchinnikov, Lee, Wang, Cong, Kinch, Schaeffer, et~al.]{baek2021accurate}
Minkyung Baek, Frank DiMaio, Ivan Anishchenko, Justas Dauparas, Sergey Ovchinnikov, Gyu~Rie Lee, Jue Wang, Qian Cong, Lisa~N Kinch, R~Dustin Schaeffer, et~al.
\newblock Accurate prediction of protein structures and interactions using a three-track neural network.
\newblock \emph{Science}, 373\penalty0 (6557):\penalty0 871--876, 2021.

\bibitem[Bose et~al.(2023)Bose, Akhound-Sadegh, Fatras, Huguet, Rector-Brooks, Liu, Nica, Korablyov, Bronstein, and Tong]{bose2023se}
Avishek~Joey Bose, Tara Akhound-Sadegh, Kilian Fatras, Guillaume Huguet, Jarrid Rector-Brooks, Cheng-Hao Liu, Andrei~Cristian Nica, Maksym Korablyov, Michael Bronstein, and Alexander Tong.
\newblock Se (3)-stochastic flow matching for protein backbone generation.
\newblock \emph{arXiv preprint arXiv:2310.02391}, 2023.

\bibitem[Darden et~al.(1993)Darden, York, and Pedersen]{darden1993particle}
Tom Darden, Darrin York, and Lee Pedersen.
\newblock Particle mesh ewald: An n log (n) method for ewald sums in large systems.
\newblock \emph{The Journal of chemical physics}, 98\penalty0 (12):\penalty0 10089--10092, 1993.

\bibitem[Dauparas et~al.(2022)Dauparas, Anishchenko, Bennett, Bai, Ragotte, Milles, Wicky, Courbet, de~Haas, Bethel, et~al.]{dauparas2022robust}
Justas Dauparas, Ivan Anishchenko, Nathaniel Bennett, Hua Bai, Robert~J Ragotte, Lukas~F Milles, Basile~IM Wicky, Alexis Courbet, Rob~J de~Haas, Neville Bethel, et~al.
\newblock Robust deep learning--based protein sequence design using proteinmpnn.
\newblock \emph{Science}, 378\penalty0 (6615):\penalty0 49--56, 2022.

\bibitem[Del~Alamo et~al.(2022)Del~Alamo, Sala, Mchaourab, and Meiler]{del2022sampling}
Diego Del~Alamo, Davide Sala, Hassane~S Mchaourab, and Jens Meiler.
\newblock Sampling alternative conformational states of transporters and receptors with alphafold2.
\newblock \emph{Elife}, 11:\penalty0 e75751, 2022.

\bibitem[Eastman et~al.(2023)Eastman, Galvelis, Pel{\'a}ez, Abreu, Farr, Gallicchio, Gorenko, Henry, Hu, Huang, et~al.]{eastman2023openmm}
Peter Eastman, Raimondas Galvelis, Ra{\'u}l~P Pel{\'a}ez, Charlles~RA Abreu, Stephen~E Farr, Emilio Gallicchio, Anton Gorenko, Michael~M Henry, Frank Hu, Jing Huang, et~al.
\newblock Openmm 8: Molecular dynamics simulation with machine learning potentials.
\newblock \emph{The Journal of Physical Chemistry B}, 2023.

\bibitem[Gur et~al.(2020)Gur, Taka, Yilmaz, Kilinc, Aktas, and Golcuk]{gur2020conformational}
Mert Gur, Elhan Taka, Sema~Zeynep Yilmaz, Ceren Kilinc, Umut Aktas, and Mert Golcuk.
\newblock Conformational transition of sars-cov-2 spike glycoprotein between its closed and open states.
\newblock \emph{The Journal of chemical physics}, 153\penalty0 (7), 2020.

\bibitem[Hansmann(1997)]{hansmann1997parallel}
Ulrich~HE Hansmann.
\newblock Parallel tempering algorithm for conformational studies of biological molecules.
\newblock \emph{Chemical Physics Letters}, 281\penalty0 (1-3):\penalty0 140--150, 1997.

\bibitem[Hoffmann et~al.(2021)Hoffmann, Scherer, Hempel, Mardt, de~Silva, Husic, Klus, Wu, Kutz, Brunton, et~al.]{hoffmann2021deeptime}
Moritz Hoffmann, Martin Scherer, Tim Hempel, Andreas Mardt, Brian de~Silva, Brooke~E Husic, Stefan Klus, Hao Wu, Nathan Kutz, Steven~L Brunton, et~al.
\newblock Deeptime: a python library for machine learning dynamical models from time series data.
\newblock \emph{Machine Learning: Science and Technology}, 3\penalty0 (1):\penalty0 015009, 2021.

\bibitem[Ingraham et~al.(2023)Ingraham, Baranov, Costello, Barber, Wang, Ismail, Frappier, Lord, Ng-Thow-Hing, Van~Vlack, et~al.]{ingraham2023illuminating}
John~B Ingraham, Max Baranov, Zak Costello, Karl~W Barber, Wujie Wang, Ahmed Ismail, Vincent Frappier, Dana~M Lord, Christopher Ng-Thow-Hing, Erik~R Van~Vlack, et~al.
\newblock Illuminating protein space with a programmable generative model.
\newblock \emph{Nature}, pp.\  1--9, 2023.

\bibitem[Janson et~al.(2023)Janson, Valdes-Garcia, Heo, and Feig]{janson2023direct}
Giacomo Janson, Gilberto Valdes-Garcia, Lim Heo, and Michael Feig.
\newblock Direct generation of protein conformational ensembles via machine learning.
\newblock \emph{Nature Communications}, 14\penalty0 (1):\penalty0 774, 2023.

\bibitem[Jing et~al.(2023)Jing, Erives, Pao-Huang, Corso, Berger, and Jaakkola]{jing2023eigenfold}
Bowen Jing, Ezra Erives, Peter Pao-Huang, Gabriele Corso, Bonnie Berger, and Tommi Jaakkola.
\newblock Eigenfold: Generative protein structure prediction with diffusion models.
\newblock \emph{arXiv preprint arXiv:2304.02198}, 2023.

\bibitem[Jorgensen et~al.(1983)Jorgensen, Chandrasekhar, Madura, Impey, and Klein]{jorgensen1983comparison}
William~L Jorgensen, Jayaraman Chandrasekhar, Jeffry~D Madura, Roger~W Impey, and Michael~L Klein.
\newblock Comparison of simple potential functions for simulating liquid water.
\newblock \emph{The Journal of chemical physics}, 79\penalty0 (2):\penalty0 926--935, 1983.

\bibitem[Jumper et~al.(2021)Jumper, Evans, Pritzel, Green, Figurnov, Ronneberger, Tunyasuvunakool, Bates, {\v{Z}}{\'\i}dek, Potapenko, et~al.]{jumper2021highly}
John Jumper, Richard Evans, Alexander Pritzel, Tim Green, Michael Figurnov, Olaf Ronneberger, Kathryn Tunyasuvunakool, Russ Bates, Augustin {\v{Z}}{\'\i}dek, Anna Potapenko, et~al.
\newblock Highly accurate protein structure prediction with alphafold.
\newblock \emph{Nature}, 596\penalty0 (7873):\penalty0 583--589, 2021.

\bibitem[Kingma \& Ba(2014)Kingma and Ba]{kingma2014adam}
Diederik~P Kingma and Jimmy Ba.
\newblock Adam: A method for stochastic optimization.
\newblock \emph{arXiv preprint arXiv:1412.6980}, 2014.

\bibitem[Lin et~al.(2023)Lin, Akin, Rao, Hie, Zhu, Lu, Smetanin, Verkuil, Kabeli, Shmueli, et~al.]{lin2023evolutionary}
Zeming Lin, Halil Akin, Roshan Rao, Brian Hie, Zhongkai Zhu, Wenting Lu, Nikita Smetanin, Robert Verkuil, Ori Kabeli, Yaniv Shmueli, et~al.
\newblock Evolutionary-scale prediction of atomic-level protein structure with a language model.
\newblock \emph{Science}, 379\penalty0 (6637):\penalty0 1123--1130, 2023.

\bibitem[Lindorff-Larsen et~al.(2011)Lindorff-Larsen, Piana, Dror, and Shaw]{lindorff2011fast}
Kresten Lindorff-Larsen, Stefano Piana, Ron~O Dror, and David~E Shaw.
\newblock How fast-folding proteins fold.
\newblock \emph{Science}, 334\penalty0 (6055):\penalty0 517--520, 2011.

\bibitem[Liu \& Nocedal(1989)Liu and Nocedal]{liu1989limited}
Dong~C Liu and Jorge Nocedal.
\newblock On the limited memory bfgs method for large scale optimization.
\newblock \emph{Mathematical programming}, 45\penalty0 (1-3):\penalty0 503--528, 1989.

\bibitem[Lu et~al.(2024)Lu, Zhong, Zhang, and Tang]{lu2024str2str}
Jiarui Lu, Bozitao Zhong, Zuobai Zhang, and Jian Tang.
\newblock Str2str: A score-based framework for zero-shot protein conformation sampling.
\newblock In \emph{The Twelfth International Conference on Learning Representations}, 2024.

\bibitem[Maier et~al.(2015)Maier, Martinez, Kasavajhala, Wickstrom, Hauser, and Simmerling]{maier2015ff14sb}
James~A Maier, Carmenza Martinez, Koushik Kasavajhala, Lauren Wickstrom, Kevin~E Hauser, and Carlos Simmerling.
\newblock ff14sb: improving the accuracy of protein side chain and backbone parameters from ff99sb.
\newblock \emph{Journal of chemical theory and computation}, 11\penalty0 (8):\penalty0 3696--3713, 2015.

\bibitem[Naritomi \& Fuchigami(2011)Naritomi and Fuchigami]{naritomi2011slow}
Yusuke Naritomi and Sotaro Fuchigami.
\newblock Slow dynamics in protein fluctuations revealed by time-structure based independent component analysis: the case of domain motions.
\newblock \emph{The Journal of chemical physics}, 134\penalty0 (6), 2011.

\bibitem[No{\'e} et~al.(2019)No{\'e}, Olsson, K{\"o}hler, and Wu]{noe2019boltzmann}
Frank No{\'e}, Simon Olsson, Jonas K{\"o}hler, and Hao Wu.
\newblock Boltzmann generators: Sampling equilibrium states of many-body systems with deep learning.
\newblock \emph{Science}, 365\penalty0 (6457):\penalty0 eaaw1147, 2019.

\bibitem[P{\'e}rez-Hern{\'a}ndez et~al.(2013)P{\'e}rez-Hern{\'a}ndez, Paul, Giorgino, De~Fabritiis, and No{\'e}]{perez2013identification}
Guillermo P{\'e}rez-Hern{\'a}ndez, Fabian Paul, Toni Giorgino, Gianni De~Fabritiis, and Frank No{\'e}.
\newblock Identification of slow molecular order parameters for markov model construction.
\newblock \emph{The Journal of chemical physics}, 139\penalty0 (1), 2013.

\bibitem[Shi et~al.(2022)Shi, Wang, Lu, Zhong, and Tang]{shi2022protein}
Chence Shi, Chuanrui Wang, Jiarui Lu, Bozitao Zhong, and Jian Tang.
\newblock Protein sequence and structure co-design with equivariant translation.
\newblock \emph{arXiv preprint arXiv:2210.08761}, 2022.

\bibitem[Song et~al.(2020)Song, Sohl-Dickstein, Kingma, Kumar, Ermon, and Poole]{song2020score}
Yang Song, Jascha Sohl-Dickstein, Diederik~P Kingma, Abhishek Kumar, Stefano Ermon, and Ben Poole.
\newblock Score-based generative modeling through stochastic differential equations.
\newblock \emph{arXiv preprint arXiv:2011.13456}, 2020.

\bibitem[Sugita \& Okamoto(1999)Sugita and Okamoto]{sugita1999replica}
Yuji Sugita and Yuko Okamoto.
\newblock Replica-exchange molecular dynamics method for protein folding.
\newblock \emph{Chemical physics letters}, 314\penalty0 (1-2):\penalty0 141--151, 1999.

\bibitem[Swendsen \& Wang(1986)Swendsen and Wang]{swendsen1986replica}
Robert~H Swendsen and Jian-Sheng Wang.
\newblock Replica monte carlo simulation of spin-glasses.
\newblock \emph{Physical review letters}, 57\penalty0 (21):\penalty0 2607, 1986.

\bibitem[Torrie \& Valleau(1977)Torrie and Valleau]{torrie1977nonphysical}
Glenn~M Torrie and John~P Valleau.
\newblock Nonphysical sampling distributions in monte carlo free-energy estimation: Umbrella sampling.
\newblock \emph{Journal of Computational Physics}, 23\penalty0 (2):\penalty0 187--199, 1977.

\bibitem[Trippe et~al.(2022)Trippe, Yim, Tischer, Broderick, Baker, Barzilay, and Jaakkola]{trippe2022diffusion}
Brian~L Trippe, Jason Yim, Doug Tischer, Tamara Broderick, David Baker, Regina Barzilay, and Tommi Jaakkola.
\newblock Diffusion probabilistic modeling of protein backbones in 3d for the motif-scaffolding problem.
\newblock \emph{arXiv preprint arXiv:2206.04119}, 2022.

\bibitem[Vani et~al.(2023)Vani, Aranganathan, Wang, and Tiwary]{vani2023alphafold2}
Bodhi~P Vani, Akashnathan Aranganathan, Dedi Wang, and Pratyush Tiwary.
\newblock Alphafold2-rave: From sequence to boltzmann ranking.
\newblock \emph{Journal of Chemical Theory and Computation}, 2023.

\bibitem[Verlet(1967)]{verlet1967computer}
Loup Verlet.
\newblock Computer" experiments" on classical fluids. i. thermodynamical properties of lennard-jones molecules.
\newblock \emph{Physical review}, 159\penalty0 (1):\penalty0 98, 1967.

\bibitem[Watson et~al.(2023)Watson, Juergens, Bennett, Trippe, Yim, Eisenach, Ahern, Borst, Ragotte, Milles, et~al.]{watson2023novo}
Joseph~L Watson, David Juergens, Nathaniel~R Bennett, Brian~L Trippe, Jason Yim, Helen~E Eisenach, Woody Ahern, Andrew~J Borst, Robert~J Ragotte, Lukas~F Milles, et~al.
\newblock De novo design of protein structure and function with rfdiffusion.
\newblock \emph{Nature}, 620\penalty0 (7976):\penalty0 1089--1100, 2023.

\bibitem[Wu et~al.(2022)Wu, Yang, Berg, Zou, Lu, and Amini]{wu2022protein}
Kevin~E Wu, Kevin~K Yang, Rianne van~den Berg, James~Y Zou, Alex~X Lu, and Ava~P Amini.
\newblock Protein structure generation via folding diffusion.
\newblock \emph{arXiv preprint arXiv:2209.15611}, 2022.

\bibitem[Yim et~al.(2023)Yim, Trippe, De~Bortoli, Mathieu, Doucet, Barzilay, and Jaakkola]{yim2023se}
Jason Yim, Brian~L Trippe, Valentin De~Bortoli, Emile Mathieu, Arnaud Doucet, Regina Barzilay, and Tommi Jaakkola.
\newblock Se (3) diffusion model with application to protein backbone generation.
\newblock \emph{arXiv preprint arXiv:2302.02277}, 2023.

\bibitem[Zhang et~al.(2019)Zhang, Liu, Yan, Tuckerman, and Liu]{zhang2019unified}
Zhijun Zhang, Xinzijian Liu, Kangyu Yan, Mark~E Tuckerman, and Jian Liu.
\newblock Unified efficient thermostat scheme for the canonical ensemble with holonomic or isokinetic constraints via molecular dynamics.
\newblock \emph{The Journal of Physical Chemistry A}, 123\penalty0 (28):\penalty0 6056--6079, 2019.

\bibitem[Zheng et~al.(2023)Zheng, He, Liu, Shi, Lu, Feng, Ju, Wang, Zhu, Min, et~al.]{zheng2023towards}
Shuxin Zheng, Jiyan He, Chang Liu, Yu~Shi, Ziheng Lu, Weitao Feng, Fusong Ju, Jiaxi Wang, Jianwei Zhu, Yaosen Min, et~al.
\newblock Towards predicting equilibrium distributions for molecular systems with deep learning.
\newblock \emph{arXiv preprint arXiv:2306.05445}, 2023.

\end{thebibliography}
\bibliographystyle{iclr2024_conference}

\newpage
\appendix

\section{Experimental details}

\subsection{Simulation protocol}\label{app:md}
The short molecular dynamics (MD) simulations were respectively initiated from each sample from the diffusion sampler. The sampled protein models are firstly pre-processed by the PDBFixer of OpenMM~\citep{eastman2023openmm} before simulation. Hydrogen atoms are added under neutral environment pH=7.0. All simulations were run on the NVIDIA Tesla V100-SXM2-32GB GPU using the Amber ff14SB force field~\citep{maier2015ff14sb} and the TIP3P water model~\citep{jorgensen1983comparison} compatible with the corresponding Amber 14 force field. $\text{Na}^+$ and $\textrm{Cl}^-$ ions were added to achieve the electrical neutrality of the solvent. The particle mesh Ewald (PME) method~\citep{darden1993particle} was used to evaluate the long-range electrostatic interaction. Each system was locally energy minimized using the L-BFGS algorithm~\citep{liu1989limited} followed by equilibration in the NVT (canonical) ensemble for 1 ns. Langevin Middle Integrator~\citep{zhang2019unified} with a friction coefficient of 1 $\textrm{ps}^{-1}$ was adopted as the thermostat (same below). Simulation temperatures for each benchmark target follows the original setting in \citet{lindorff2011fast}, specifically, unit in Kelvin (K): Chignolin: 340K, Trp-cage: 290K, BBA: 325K, Villin: 360K, WW domain: 360K, NTL9: 355K, BBL: 298K, Protein B: 340K, Homeodomain: 360K, Protein G: 350K, $\alpha$3D: 370K and lambda-repressor: 350K. Then the system was equilibrated in the NPT (isothermal-isobaric) ensemble for 1 ns while coupled with the Monte Carlo barostat at constant 1 bar pressure. All production simulations were performed using OpenMM v8.0.0~\citep{eastman2023openmm} in the NPT ensemble integrated with time step of $2.5$ fs and trajectory frames were saved per 10 ps of simulation.

\subsection{Fine-tuning dataset}
To acquire the fine-tuning dataset (ES) for each target, we firstly generate the seeding structure using ESMFold~\citep{lin2023evolutionary} with default configuration and ran Str2Str as described in \citet{lu2024str2str} to generate 100 conformations from the seeding structure. Then MD simulations was performed in parallel initialized respectively from these conformations according to the protocol above in Appendix \ref{app:md}. We ignored those simulation threads failed due to possible bad initial geometries.  The resulting conformations as well as their associated potential energies were saved to construct the fine-tuning dataset. We randomly split the dataset into training and validation set with ratio $0.95:0.05$ while the latter is used for early stopping of training.

\subsection{Setup of training and inference}
To optimize the network parameters of the pre-trained Str2Str, the Adam optimizer~\citep{kingma2014adam} was used with $\rm{lr}= 10^{-4}$ and $\beta_1=0.9, \beta_2=0.999$. As scheduler, the learning rate was reduced in the factor of 0.1 when the loss has stopped decreasing for 10 epochs. The training  process was set to be at least 100 epochs and mediated the early stopping strategy by monitoring the validation loss with 10 epochs patience. The fine-tuning time for different targets was on-average $\sim$ 2 GPU hours on the NVIDIA Tesla V100-SXM2-32GB GPU. For simplicity, we use uniform weighting in this study for the training objective, i.e. $\omega(x, t)\equiv 1$. Other model configurations and evaluation pipelines were kept the same as in \citet{lu2024str2str}. After fine-tuning, we sampled the target protein from the prior distribution $p_T(\rvx)$ with discretization of 1,000 timesteps by reversing the diffusion process via Langevin dynamics. In practice, a minimum time $t_\epsilon=0.01$ instead of zero is used for numerical stability. The SDE mode of sampling was used for both Str2Str and Str2Str-FT.

\subsection{Fast-folding Targets} The fast-folding targets with the reference MD trajectories are listed as follows~\citep{lindorff2011fast}: Chignolin, Trp-cage, BBA, Villin, WW domain, NTL9, BBL, Protein B, Homeodomain, Protein G, $\alpha$3D and Lambda-repressor. The reference MD trajectories for evaluation were obtained by sending requests to the authors of \citet{lindorff2011fast}.

\section{Evaluation metrics}\label{app:eval}
In this section, we elaborate the definition of the evaluation metrics. For validity, the Val-Clash is defined by the ratio of samples in the predicted ensemble of size 1,000 which do not contain any clash. Clash is detected by examining whether there exists pairwise distance between C$\alpha$ atoms less than a distance threshold of $\delta_c=3.0$\AA; the Val-Bond is defined similarly as the ratio of samples having no (pseudo) bond dissociation. The dissociation is counted if there is any distance between adjacent C$\alpha$ atoms that exceeds certain threshold $\delta_b$, where $\delta_b$ is defined by the maximum value of distances between adjacent C$\alpha$ in the reference MD trajectory of the target system. For distribution divergence metrics, we calculate three features including pairwise distance (PwD), time-lagged independent component (TIC) coordinates, and radius of gyration (Rg) for each conformation in the predict ensemble. To transform the continuous feature values into distribution, histograms are built with $N_{\rm bin} = 50$ bins to represent the categorized distribution over which the JS divergence is calculated. For each channel of histogram, a pseudo-count value of $\epsilon = 10^{-6}$ was added to offset zero frequencies and slightly smooth the distribution. The pairwise distance is computed by enumerating all pairs of C$\alpha$ atoms with an offset three; the TICA dimension reduction is performed using the Deeptime library~\citep{hoffmann2021deeptime} and the first two slowest dimensions are selected for evaluation; and the radius of gyration is defined by the root mean square distance of each C$\alpha$ atoms relative to the center of mass of the protein. All of three distributions are compared respective to the reference long MD simulation to calculate the Jensen-Shannon (JS) divergence metrics.

% Note that we omit the computationally extensive \textit{diversity} metrics in \citet{lu2024str2str} which involve pairwise TM-score since the distribution distance metrics (JS) can also reflect diversity mismatch.

\section{Extended experimental results}
\paragraph{Compare neighborhood exploitation with MD simulation}
We demonstrate the effectiveness of the neighborhood exploitation (NE) over conventional MD simulation in Table \ref{tb:abl_md}. The exploration over conformation space of MD simulation (under limited budget) can be strongly enhanced by firstly sampling from a pre-trained diffusion sampler. The simulation protocol of both Str2Str-NE and MD are kept the same as described in \ref{app:md}.

\begin{table}[ht!]
    \centering
    % \scriptsize
    % \small
    \caption{Comparison between samples obtained from neighborhood exploitation (parallel short simulations) and computation-equivalent conventional MD simulation.}
    \label{tb:abl_md}
    \begin{tabular}{c|ccccc}%{c|c|c|c|c|c|c}
    \toprule[1pt]
        Setting  & Val-Clash($\uparrow$) & Val-Bond($\uparrow$) & {JS-PwD}($\downarrow$) & {JS-TIC}($\downarrow$) &  JS-Rg ($\downarrow$) \\ 
        \midrule
        MD (100 ns)  & \textbf{1.000} & \textbf{1.000} & 0.518 & 0.547 & 0.523  \\
        Str2Str-NE & 0.990 & \textbf{1.000} & \textbf{0.294}  & \textbf{0.369}& \textbf{0.331} \\
        \bottomrule[1pt]
    \end{tabular}
\end{table}

\begin{figure}[ht!]
    \centering
    \includegraphics[width=0.9\linewidth]{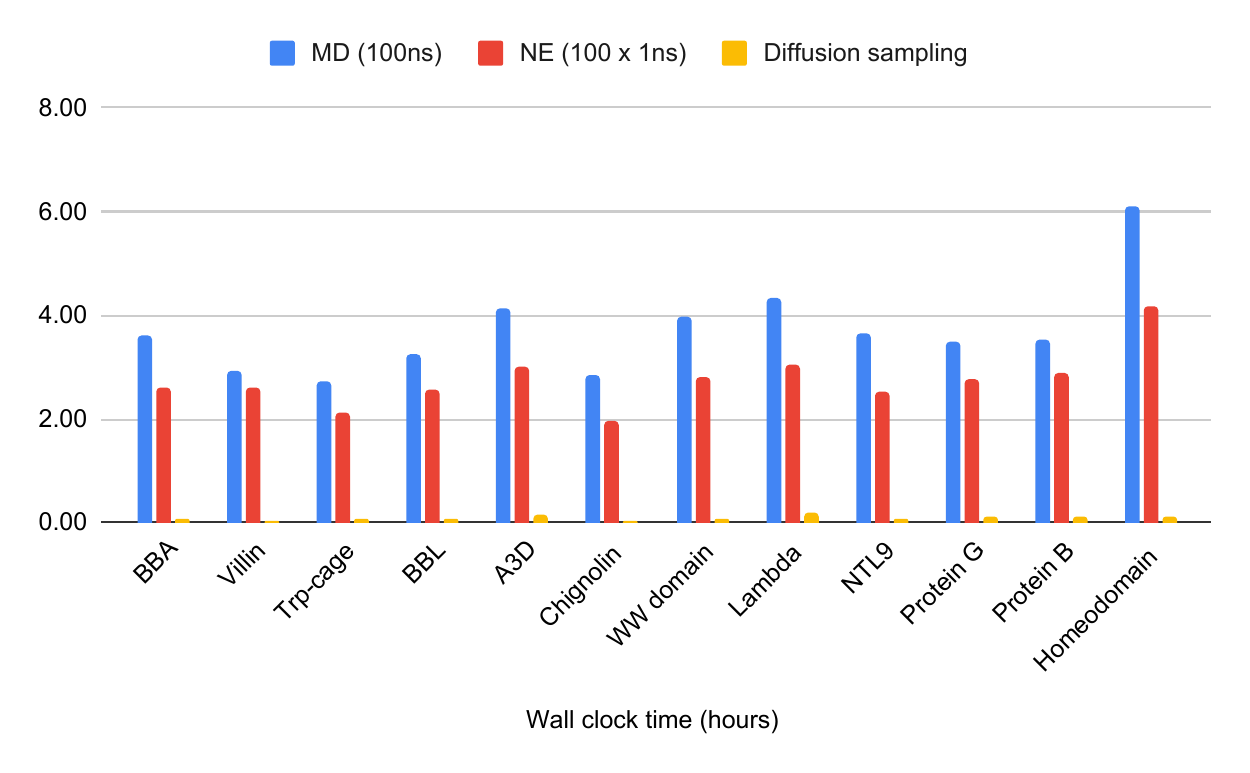}
    \caption{The runtime profile of different samplers across each fast folding target used in this study. }
    \label{fig:runtime}
\end{figure}

\paragraph{Runtime Analysis}
To illustrate the efficiency of proposed few-shot sampling over conventional MD simulation, we here profile the \textit{wall clock} running time of each type of "sampler" for each benchmark target, shown in Table \ref{fig:runtime}. Note that the runtime can vary from different proteins due to the different number of atoms. The profiling results are all reported as the wall clock hours based on 4$\times$ NVIDIA Tesla V100-SXM2-32GB GPUs. 
% The efficiency of NE over conventional MD is that MD has very bad scalability with increasing number of GPUs due to the overhead of communications; in contrast, the short simulations conducted in NE are independent and thus scale well with the increasing computational resources.

% 100ns overhead: communication 
% es overhead: simulation  preparation (minimize and equilibration)

% \input{tables/runtime}

\section{Limitations and Discussion}
Similar to Str2Str~\citep{lu2024str2str}, the proposed methods cannot guarantee the sampled ensemble are Boltzmann distributed~\citep{noe2019boltzmann} due to the nature of local simulation. It is not designed for replacing canonical MD simulations for the study of protein dynamics. Incorporating short MD simulations in the Str2Str sampler takes the advantages from both physical and neural sides: the neural sampler (eg., diffusion) is good at proposing data-like conformation hypothesis very efficiently in the vast conformation space while the physical simulation can equilibrate the samples onto some local minima. Both neighborhood exploitation (NE) and fine-tuning (FT) provide a simple yet effective way to refine the resulting conformations sampled from a neural sampler. 

From another perspective, neighborhood exploitation coincides with the purpose of enhanced sampling methods. Enhanced sampling methods aim to overcome the energy barrier and accelerate the exploration of long MD simulations. For example, the umbrella sampling~\citep{torrie1977nonphysical}, and parallel tempering (synonymous replica exchange molecular dynamics, REMD)~\citep{hansmann1997parallel, sugita1999replica, swendsen1986replica}. The neural sampler such as Str2Str can provide diverse conformation seeds to augment the exploration of MD simulation similar to \citet{vani2023alphafold2}. We find it a promising research direction to develop the neural enhanced sampling methods in the future.

\end{document}